\documentclass[twocolumn,prl,tightenlines,superscriptaddress,showpacs]{revtex4-2}

\usepackage{amsmath}
\usepackage{amssymb,amsfonts,latexsym}
\usepackage{bm}
\usepackage[mathcal]{euscript}
\usepackage{graphicx}
\usepackage{soul}

\usepackage{xcolor}
\usepackage{siunitx}

\newcommand{\modif}[1]{\textcolor{black}{#1}}

\begin{document}

\title{Giant density fluctuations in locally hyperuniform states}

\author{Sara Dal Cengio}
\email{saradc@mit.edu}
\affiliation{Department of Physics, Massachusetts Institute of Technology, Cambridge, Massachusetts 02139,
USA}
\affiliation{Universit\'e Grenoble Alpes, CNRS, LIPhy, 38000 Grenoble, France}

\author{Romain Mari}
\affiliation{Universit\'e Grenoble Alpes, CNRS, LIPhy, 38000 Grenoble, France}

\author{Eric Bertin}
\affiliation{Universit\'e Grenoble Alpes, CNRS, LIPhy, 38000 Grenoble, France}

\date{\today}

\begin{abstract}
Systems driven far from equilibrium may exhibit anomalous density fluctuations: active matter with orientational order
display giant density fluctuations at large scale, while systems of interacting particles close to an absorbing phase transition may exhibit hyperuniformity,
suppressing large-scale density fluctuations. We show that 
these seemingly incompatible phenomena can coexist in nematically ordered active systems, provided activity is conditioned to particle contacts. We characterize this unusual state of matter and unravel the underlying mechanisms simultaneously leading to spatially enhanced (on large length scales) and suppressed (on intermediate length scales) density fluctuations. Our work highlights the potential for a rich phenomenology in active matter systems in which particles' activity is triggered by their local environment, and calls for a broader exploration of absorbing phase transitions in orientationally-ordered particle systems.
\end{abstract}

\maketitle

Outside critical points, particle systems at thermal equilibrium have a finite compressibility and display normal density fluctuations in homogeneous phases.
By contrast, athermal systems of polydisperse particles like disordered sphere packings \cite{Berthier2011,TorquatoPRL2011}, dense emulsions \cite{Ricouvier2017}, colloidal suspensions \cite{Ma2020} may display hyperuniformity, whereby large-length-scale density fluctuations are suppressed \cite{torquatoLocalDensityFluctuations2003,torquato2018,ma_torquato2019,TorquatoPRE2021}. 
For monodisperse athermal particle systems, hyperuniformity is often observed at the critical point of an absorbing phase transition
\cite{hexner2015,tjhung2015,schrenkCommunicationEvidenceNonergodicity2015,weijsEmergentHyperuniformityPeriodically2015,hexnerNoiseDiffusionHyperuniformity2017a,hexnerEnhancedHyperuniformityRandom2017,maTheoryHyperuniformityAbsorbing2023} (although not always~\cite{mariAbsorbingPhaseTransitions2022}).
Physical realizations include cyclically sheared suspensions
\cite{corteRandomOrganizationPeriodically2008,weijsEmergentHyperuniformityPeriodically2015,tjhungCriticalityCorrelatedDynamics2016,wangHyperuniformityNoFine2018},
whose stroboscopic dynamics has been schematically described by the Random Organization Model (ROM) \cite{corteRandomOrganizationPeriodically2008,milzConnectingRandomOrganization2013,tjhung2015,hexner2015}, 
as well as some systems of active particles~\cite{leiHydrodynamicsRandomorganizingHyperuniform2019a,LeiScienceAdvances2019},
or topological defects in active nematics \cite{fernandez2024hyperuniform}.
Hyperuniform regimes have also been reported in systems with long-range (e.g., hydrodynamic) interactions (like chiral active particles \cite{HuangPNAS2021}, spinners or vortices in a fluid \cite{ShelleyNatCom2022}, or microvortices in active turbulence \cite{TorquatoPNAS2024}), or
short-range antialigning interactions~\cite{boltz2024hyperuniformity}.
Finally, clustering or coarsening patterns lead to hyperuniformity on very large length scales in scalar active matter
\cite{LowenPRR2024,zhangHyperuniformActiveChiral2022,lucaHyperuniformityPhaseOrdering2024}, where no large-scale orientational order develops.

By contrast, systems of orientationally-ordered active particles are notoriously characterized by giant number fluctuations (GNF). 
Large-length-scale density fluctuations exceed the ones of an equilibrium system, due to the slow decay of the two-point density correlation function which arises from the coupling with orientational order~\cite{ramaswamy2003, matter2022walking}. GNF occur without fine tuning of parameters and are predicted to appear generically in the orientationally ordered phase of active matter.  GNF have been first predicted using fluctuating hydrodynamics~\cite{ramaswamy2003}, and afterwards observed in numerical~\cite{chate2006,Dey2012, ngo2014} and experimental realizations, including vibrated rods~\cite{narayan2007long}, vibrated polar disks \cite{Deseigne2010}, bacteria colonies \cite{Zhang2010collective}, and epithelial cells \cite{Giavazzi2017}.

It is thus of interest to quantitatively characterize the density fluctuations in a system which mixes the key ingredients to GNF and hyperuniformity. Indeed, many active particle systems can potentially exhibit orientational order (e.g., due to alignment interactions) and an absorbing phase transition (e.g., due to a motility that is activated by the presence of neighbors).
Experimentally feasible systems may include suspensions of rods-like particles under cyclic shear near their reversible-irreversible transition~\cite{franceschiniDynamicsNonBrownianFiber2014,trulssonDirectionalShearJamming2021},
Quincke rollers with neighbor-triggered motility \cite{LiuActivityPNAS2021},
assemblies of biological cells with stress-induced motility \cite{VedelPNAS2013,Putelat2018}, collections of active microtubules \cite{sumino2012large}, or swarms of macro- \cite{Chen2024Emergent} or micro-robots~\cite{muinos2021,benzion2023}.
Understanding how the interplay of interaction-triggered motility and orientational order impacts large-scale density fluctuations may thus open avenues for designing active systems with self-organized complex states.

In this Letter, we introduce an active nematics model exhibiting an absorbing phase transition similar to the one of the ROM,
that we coin the nematic ROM (NROM). \modif{This minimal model is inspired by experiments on cyclically-sheared suspensions of rods \cite{franceschiniDynamicsNonBrownianFiber2014,trulssonDirectionalShearJamming2021} and on subcritical Quincke rollers \cite{LiuActivityPNAS2021},
which both display neighbor-triggered motility and orientational ordering.}
We show that when the absorbing phase transition occurs in the nematically ordered state, the system dynamically organizes to suppress density fluctuations in an intermediate range of length scales when approaching the transition,
and to enhance fluctuations at larger length scales, where GNF are observed. 
We predict a critical crossover between these two anomalous regimes of density fluctuations, which we characterize using a continuum theory. In this framework, the crossover arises from the competition between two effective noises on the density field.
More generally, our results open avenues for future research on the detailed characterization of universality classes of absorbing phase transitions in active matter with orientational order and environment-triggered motility.

\begin{figure}
  \centering
  \includegraphics[width=\columnwidth]{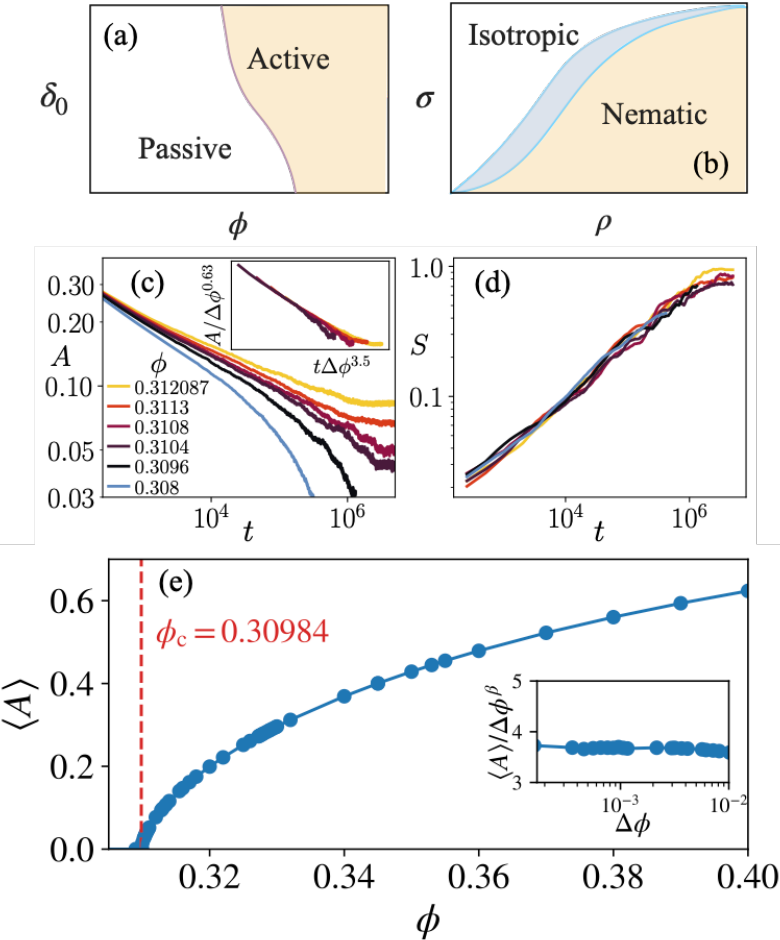}
\caption{(a) Sketch of the phase diagram for the ROM model showing the  active and passive phases in the volume fraction $\phi$ -- step size $\delta_0$ plane, inspired by Fig.~1 in \cite{tjhung2015}. (b) Sketch of the  phase diagram for active nematics showing the isotropic and  nematically ordered phases in the density $\rho$ -- noise amplitude $\sigma$ plane, inspired by Fig.~2 in \cite{ngo2014} (blue shaded area: phase-separated nematic state). The NROM model: (c) Activity $A$ (data collapse shown in the inset) and (d) scalar nematic order $S$ versus time, for several packing fractions $\phi$. Data are averaged over more than 40 realizations. (e) Steady-state activity $\langle A \rangle$ versus packing fraction $\phi$.
}
\label{fig:sketch}
\end{figure}

We consider a twodimensional system of $N$ particles with diameter $D$ in a box of linear size $L$ and periodic boundary conditions.
Each particle $i = 1, \dots, N$ at position $\bm{r}_i^{t}$ carries an axial direction $\theta^{t}_i \in [-\pi /2, \pi /2]$ and aligns noisily to the neighboring particles within the interaction range $R_{\mathrm{al}}$. At each time step, the directors $\bm{n}^{t}_i = \left( \cos \theta^{t}_i, \sin \theta^{t}_i\right)$ are updated according to
a Vicsek-type active nematics rule \cite{chate2006,ngo2014},
\begin{equation} \label{eq:align:rule}
    \theta^{t+1}_i = \frac{1}{2} \text{Arg} \left[ \sum_{k \in \mathcal{V}_i} e^{\mathrm{i} 2 \theta^{t}_{k}}\right] + \psi^{t}_i \; \mod \pi
\end{equation}
where $\mathcal{V}_i$ is the neighborhood of particle $i$ and $\psi^{t}_i \in [-\sigma \pi /2, \sigma \pi/2]$ is a random angle drawn from a uniform distribution,
with $\sigma \in [0,1]$ the noise intensity.
Unlike usual active nematics models \cite{chate2006,ngo2014} where all particles permanently diffuse, 
here only particles overlapping with another particle (i.e., with a center-to-center distance smaller than $D$) are motile.
The position of overlapping particles is updated as:
\begin{equation}
    \bm{r}_i^{t+1} = \bm{r}_i^{t} \pm \delta_0 \bm{n}^{t}_i\,,
\end{equation}
with random equiprobable signs, and $\delta_0$ a fixed step size.
Our model thus generalizes the Vicsek-type active nematics model \cite{chate2006,ngo2014} to a ROM-type model with nematic interactions and anisotropic diffusion, such that particles diffuse along their individual director $\bm{n}^{t}_i$, only when in overlap with another particle (however, all angles $\theta_i$ are updated at each time step).
The NROM boils down to the ROM for maximal angular noise ($\sigma=1$), while active nematics is recovered for $D \to \infty$ (interpreting $D$ as an interaction range for triggering motion).

We start by recalling the schematic phase diagrams of the ROM and active nematics models.
The control parameters of the ROM are the packing fraction $\phi = \frac{N \pi D^{2}}{4 L^2}$ and the step size $\delta_0$. Its phase diagram is sketched in Fig.~\ref{fig:sketch}(a): a passive phase at small $\phi$ for which the system always falls into an absorbing state after a finite number of time steps is separated from an active phase at large $\phi$ by an absorbing phase transition (purple line).
The control parameters of the Vicsek-type active nematics model \cite{chate2006,ngo2014} are the dimensionless density $\rho=\frac{N R_{\mathrm{al}}^2}{L^2}$ and the noise intensity $\sigma$ [Eq.~(\ref{eq:align:rule})].
An isotropic state is obtained for large noise and small density, while nematic order sets in for small noise and large density [Fig.~\ref{fig:sketch}(b)], with a phase-separated region close to the transition.

The combined model (NROM) has four control parameters, and therefore a potentially complex phase diagram.
However we aim at a restricted region within this phase diagram: 
close to the absorbing phase transition where hyperuniformity could arise 
(i.e., for $\phi$ around the critical packing fraction $\phi_\mathrm{c}$) 
in an otherwise nematically ordered phase showing GNF 
(i.e., for large $\rho$). 
These two conditions can be simultaneously satisfied for $D/R_{\rm al}$ smaller than a threshold value that depends on $\sigma$ and $\delta_0$ values.
Here we use $\sigma =0.1$, $\delta_0=0.3$, $R_{\rm al} = 1$, and pick  $D=0.46$~\footnote{This particular choice ensures that the absorbing phase transition occurs near $\rho=2$, for which the active nematics study of~\cite{ngo2014} was performed. Other choices are possible, see e.g. for $\sigma = 0.05$, $\delta_0=0.3$, $R_{\rm al} = 1$, and $D=0.3$~\cite{SuppMat}}. The system size is $L=516$ unless specified otherwise, and simulations are initialized with random uniformly distributed positions and angles.

We show in Fig.~\ref{fig:sketch}(c) the activity $A$, defined as the fraction of active (i.e., overlapping) particles, as a function of time.
An absorbing phase transition is observed at a packing fraction $\phi_\mathrm{c}$.
For $\phi>\phi_\mathrm{c}$, the activity reaches a steady state around an averaged value $\langle A \rangle>0$ in a time that diverges when $\phi\to\phi^+_\mathrm{c}$ (see Fig.~\ref{fig:sketch}(c) inset). 
By contrast, for $\phi<\phi_\mathrm{c}$, the activity dies out in a finite time, and particle positions become frozen.
In Fig.~\ref{fig:sketch}(e), we show $\langle A\rangle$ as a function of $\phi$ and the corresponding critical behavior $\langle A \rangle/\Delta\phi^\beta$ versus $\Delta\phi = \phi-\phi_\mathrm{c}$, from which we estimate $\phi_\mathrm{c} = 0.30984 \pm 0.00006$.
A fit yields an exponent $\beta \approx 0.63\pm 0.01$ [Fig.~\ref{fig:sketch}(e)], a value compatible with the twodimensional conserved directed percolation (CDP) class \cite{Lubeck2004}, the university class of absorbing phase transition to which the ROM belongs \cite{Menon2009, tjhungCriticalityCorrelatedDynamics2016, bubleeHyperuniformityAbsorbingPhase2019}.
Besides, the traceless nematic tensor is defined as
$\bm{Q} = N^{-1}\sum_i \bm{n}_i^t \bm{n}_i^t -\frac{1}{2}\bm{1}$, with $\bm{1}$ the identity matrix, and
the nematic scalar order parameter reads $S = 2 \sqrt{|\det{\bm{Q}}|}$ ($0 \leq S \leq 1$).
For $\phi>\phi_\mathrm{c}$, the system reaches a highly ordered nematic state with $S > 0.9$ [Fig.~\ref{fig:sketch}(d)] within a time which does not diverge at $\phi_\mathrm{c}$
(for $\phi<\phi_\mathrm{c}$, the simulation is stopped when falling into an absorbing state). In the following, we focus on the case $\phi > \phi_\mathrm{c}$.

\begin{figure}[t]
\centering
\includegraphics[width=\columnwidth]{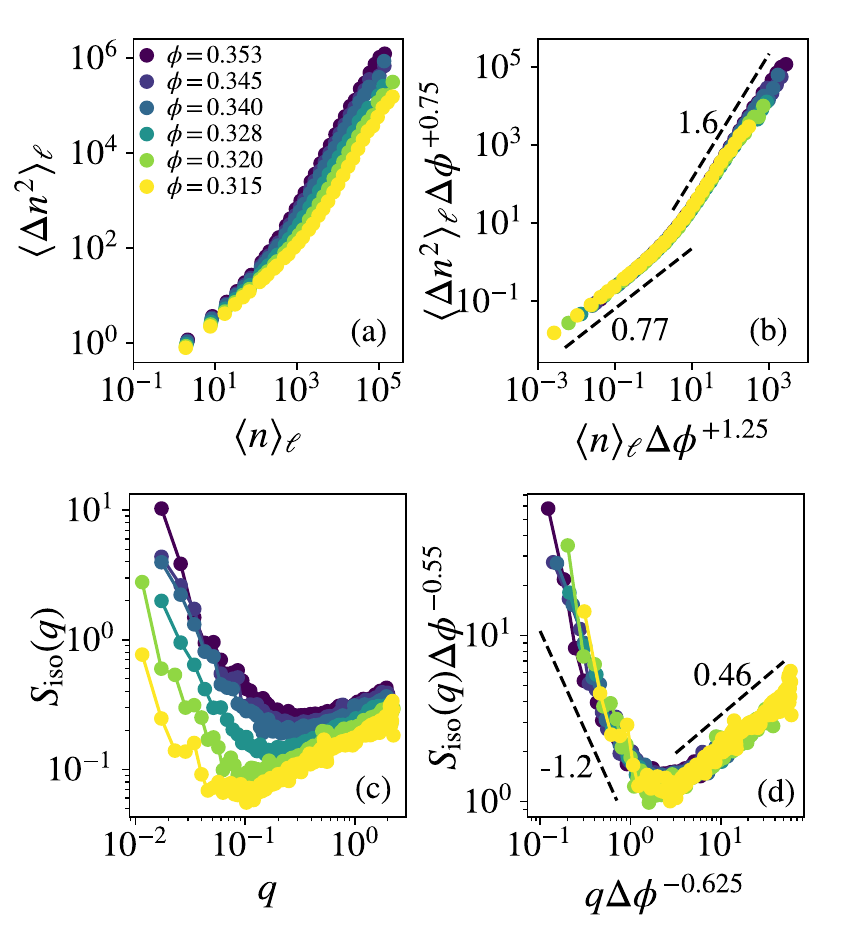}
\caption{
(a) Number fluctuations $\langle \Delta n^2  \rangle_{\ell}= \langle n^2 \rangle_{\ell} - \langle n \rangle_{\ell}^2$ versus $\langle n \rangle_{\ell}$ in logarithmic scale for several packing fractions $\phi > \phi_\mathrm{c} = 0.30984$.  (b) Same data, rescaled as $\langle \Delta n^2 \rangle_{\ell} / \Delta \phi^{-0.75}$ versus $\langle n \rangle_{\ell}/\Delta \phi^{-1.25}$.  (c) Structure factor $S(q)$ for several values of $\phi > \phi_\mathrm{c}$ in logarithmic scale. (d) Same data, rescaled as $S(q)/\Delta \phi^{0.55}$ versus $q/\Delta \phi^{0.625}$. 
For $\phi =0.353, 0.345, 0.340, 0.328$ system size $L=516$ and  for $\phi =  0.320, 0.315$ system size $L=774$.}
\label{fig:figure2}
\end{figure}

To characterize density fluctuations, we evaluate the variance $\langle \Delta n^2 \rangle_{\ell} = \langle n^2 \rangle_{\ell} - \langle n \rangle_{\ell}^2$ of the particle number $n$ within boxes of different linear size $\ell$. 
We plot in Fig.~\ref{fig:figure2}(a) the variance $\langle \Delta n^2 \rangle_{\ell}$ versus $\langle n \rangle_{\ell}$, parameterized by the box size $\ell$,
for different $\phi$ close to the critical point.
Normal fluctuations follow a power law $\langle \Delta n^2 \rangle_{\ell} \sim \langle n \rangle_{\ell}^{\alpha}$ with $\alpha=1$, 
while $\alpha > 1$ corresponds to GNF~\cite{ramaswamy2003,chate2006,ngo2014}, and
$\alpha<1$ corresponds to hyperuniformity \cite{hexner2015,torquato2018}.
We identify two distinct regimes of anomalous density fluctuations: at intermediate length scales the system develops hyperuniformity ($\alpha = \alpha_{\mathrm{hu}} = 0.77 \pm 0.005$, consistently with the value found in the ROM \cite{hexner2015}), while for large length scales the system displays GNF ($\alpha = \alpha_{\mathrm{gnf}} = 1.65 \pm 0.05$, as found in the $2D$ nematic Vicsek model \cite{ngo2014}), see Fig.~\ref{fig:figure2}(a).
The crossover length $\xi$ between the two regimes grows when approaching the critical point ($\phi\to \phi_{\mathrm{c}}^{+}$).
We assume a power-law scaling $\xi \sim \Delta \phi^{-\mu}$ (with $\mu$ to be determined),
corresponding to a characteristic number of particles  $n^* = \rho \xi^2 \sim \Delta \phi^{-2\mu}$.
This critical scaling suggests a possible data collapse by plotting $\langle \Delta n^2 \rangle_{\ell} \Delta \phi^{\zeta}$ versus $\langle n \rangle_{\ell} \Delta \phi^{2\mu}$,
under appropriate choices of the exponents $\mu$ and $\zeta$.
We find a good data collapse for $\mu=0.625 \pm 0.02$ and $\zeta=0.75 \pm 0.05$ [Fig.~\ref{fig:figure2}(b)]. The divergence of the crossover length $\xi$ at $\phi_\mathrm{c}$ tells that by getting close enough to the critical point one may observe hyperuniformity on arbitrarily large length scales.

\modif{Density fluctuations are usually also  characterized with 
the structure factor $S(\bm{q}) = N^{-1} \langle \sum_{i, j} e^{i \bm{q}\cdot(\bm{r}_i - \bm{r}_j)} \rangle$,
which is experimentally accessible for systems of interest~\cite{wilkenHyperuniformStructuresFormed2020}.}
Since $S(\bm{q})$ may be anisotropic due to the nematic order, we introduce the angular averaged structure factor $S_\mathrm{iso}(q)$ with $q = |\bm{q}|$. 
For small $q\ll 2\pi/D$, $S_\mathrm{iso}(q)$ maps to $\langle \Delta n^2 \rangle_{\ell}$ \cite{torquato2018}:
\begin{equation}\label{eq:scalingSqDeltan}
    S_\mathrm{iso}(q) \sim \frac{\langle \Delta n^2 \rangle_{\ell}}{\langle n \rangle_{\ell}} \quad  \quad \text{with} \quad q \sim \frac{2\pi}{\ell} \, .
\end{equation}
If $\langle \Delta n^2 \rangle_{\ell} \sim \langle n \rangle_{\ell}^{\alpha}$, then $S_\mathrm{iso}(q) \sim q^{\lambda}$ for small $q$,
with $\lambda = (1 -\alpha) d$ \cite{torquato2018}, with $d$ the space dimension.
For the 2D Manna model (a lattice sandpile model in the CDP class~\cite{mannaTwostateModelSelforganized1991}), this relation gives $\lambda_{\rm hu}\approx 0.46$, corresponding to hyperuniformity~\cite{hexnerNoiseDiffusionHyperuniformity2017a}.
In contrast, for 2D active nematics this gives $\lambda_{\rm gnf}\approx  -1.2$, leading to GNF.
We plot on Fig.~\ref{fig:figure2}(c) the structure factor evaluated numerically for the same values of $\phi$ as in Fig.~\ref{fig:figure2}(a).
We again find two scaling regimes.
For very small $q$, we observe a divergence compatible with a power law $q^{\lambda_{\mathrm{gnf}}}$, with $\lambda_{\mathrm{gnf}}<0$.
For intermediate values of $q$, we rather observe a power law $q^{\lambda_{\mathrm{hu}}}$ with $\lambda_{\mathrm{hu}}>0$.
We obtain a data collapse under an appropriate rescaling [Fig.~\ref{fig:figure2}(d)], by plotting $S_\mathrm{iso}(q)/ \Delta\phi^{\zeta'}$ versus $q /\Delta\phi^{\mu}$ with
$\zeta'= 0.55 \pm 0.05$ and $\mu= 0.625 \pm 0.05$ (same value of $\mu$ as in Fig.~\ref{fig:figure2}(b)), thereby confirming the critical crossover between hyperuniformity and GNF around a wavenumber $q^* \sim \xi^{-1}\sim \Delta\phi^{\mu}$.
Using $\langle n \rangle_{\ell} \sim \ell^{2} \sim q^{-2}$, Eq.~(\ref{eq:scalingSqDeltan}) yields $\zeta' = 2 \mu - \zeta \approx  0.5$, which is compatible with our numerical estimate. The behavior of the structure factor, with giant density fluctuations on length scales up to system size, highlights the role of nematic order. 
By contrast, for $\phi<\phi_\mathrm{c}$, where nematic ordering does not have time to build up, a crossover to normal fluctuations is observed at $q\to 0$~\cite{SuppMat}, just like in the usual ROM~\cite{tjhungCriticalityCorrelatedDynamics2016,hexnerEnhancedHyperuniformityRandom2017}.

\begin{figure}[t]
\centering
\includegraphics[width=\columnwidth]
{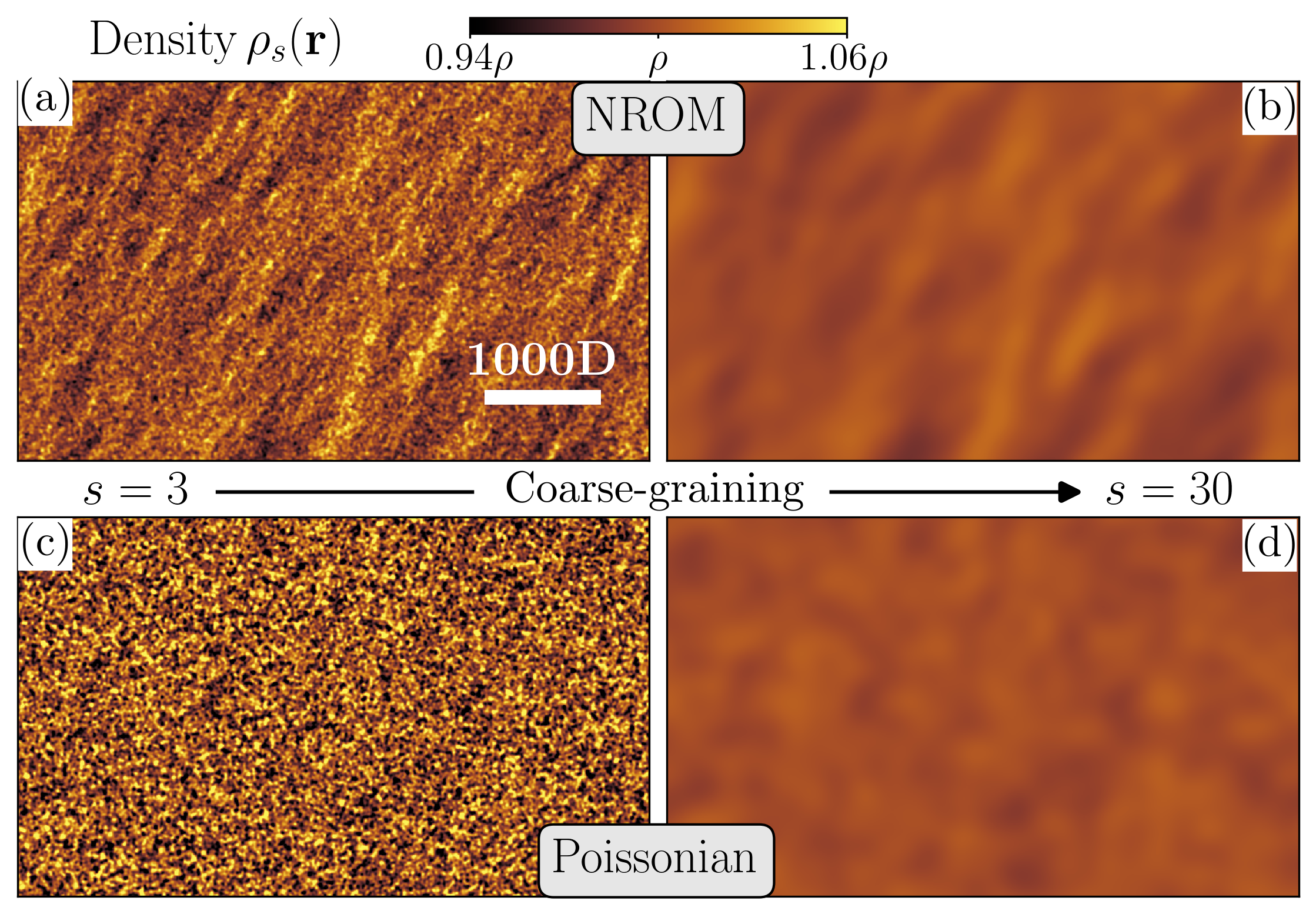}
\caption{Density fields of steady-state configurations from the NROM model [top row, (a)--(b)] and of a Poissonian realization [bottom row (c)--(d)] with same 
system size $L = 2850$ and 
packing fraction $\phi =0.34327$, obtained with two different values of the coarse-graining length $s$: $s=3$ [(a) and (c)] and $s=30$ [(b) and (d)].}\label{fig:snaps}
\end{figure}

To illustrate the coexistence of different density fluctuation regimes, we compare in Fig.~\ref{fig:snaps} the density fields $\rho_s(\bm{r}) = (2\pi s^2)^{-1}\sum_i \exp[-(\bm{r}-\bm{r}_i)^2/2s^2]$ obtained from the same steady-state particle configuration under two values of the coarse-graining length $s$ [Fig.~\ref{fig:snaps}(a,b)] to a random (Poissonian) particle configuration [Fig.~\ref{fig:snaps}(c,d)]. 
Compared to the random configuration, the NROM density field appears comparably smoother (more uniform) than the random configuration for small $s$, illustrating hyperuniformity, whereas it shows fluctuations of larger amplitude and length scales for large $s$ that are consequences of GNF.

To get some insight on the crossover length $\xi$,
we introduce a hydrodynamic description based on the three relevant physical fields: 
the particle density $\rho(\bm{r}, t)$, the nematic 
tensor field $\bm{Q}(\bm{r}, t)$, and the activity field $A(\bm{r},t)$ defined as the density of overlapping particles. 
Particles move only when they are active, so the same particle current $\bm{J}$ appears in the evolution equations for $\rho$ and $A$.
In a gradient expansion, the most general deterministic part of the particle flux is 
$\bm{J} = D_\rho \nabla A + \chi_1 \bm{Q}\cdot \nabla A + \chi_2 A\nabla\cdot \bm{Q}$, where $D_\rho>0$, $\chi_1$ and $\chi_2$ are parameters. The last two terms represent curvature-driven currents, specific of active nematics.
The continuity equation for the (conserved) density field reads
\begin{equation}\label{eq:densityevolution}
    \partial_t \rho = \nabla \cdot \left( \bm{J} + \sigma_\rho \sqrt{A} \bm{\eta}_{\rho} \right)
\end{equation}
with $\sigma_\rho$ a positive parameter and $\bm{\eta}_{\rho}(\bm{r}, t)$ a vectorial unit Gaussian white noise, $\langle \bm{\eta}_{\rho} (\bm{r}, t) \bm{\eta}_{\rho} (\bm{r}', t') \rangle = \delta(\bm{r} - \bm{r}')\delta(t - t') \bm{1}$.
We assume the same reaction terms for the activity as in the CDP universality class \cite{Pastorsatorras2000, Menon2009}, so that
\begin{equation}\label{eq:activityevolution}
    \partial_t A = \nabla \cdot \bm{J} + (\kappa \rho - a)  A - \lambda A^2 + \sigma_A \sqrt{A} \eta_{A}
\end{equation}
where $\kappa$, $a$, $\lambda$, $\sigma_A$, are positive parameters, and $\eta_{A}(\bm{r}, t)$ is a scalar unit Gaussian white noise. The reaction terms in Eq.~(\ref{eq:activityevolution}) accounts for the fact that the activity field $A$ is not conserved: upon a random displacement, an active particle may become passive, or may activate a passive particle by overlapping with it. Moreover, $A$ is nonzero in stationary state only for $\rho>\rho_\mathrm{c}$.
The noise on particle current is neglected with respect to the nonconserved noise.
The effect of the nematic field $\bm{Q}$ is limited to the transport term $\nabla\cdot \bm{J}$, and is the only difference with the usual CDP dynamics for $A$.
To lighten the presentation, we set here $\chi_1 = 0$ in $\bm{J}$, which has no qualitative effect on the density fluctuations crossover
(the full calculation with $\chi_1 \ne 0$ is given in \cite{SuppMat}).
Particles align with their neighborhood irrespective of their activity, but nematic order is transported only by diffusion of active particles, yielding the following evolution for the nematic tensor
\begin{equation}\label{eq:nematicevolution}
        \partial_t \bm{Q} = \left(\tilde{\mu}(\rho) - \gamma |\bm{Q}|^{2}\right) \bm{Q} + D_Q \nabla^2 \bm{Q} + \chi_3 \widehat{\nabla \nabla}  A  + \sigma_Q \bm{\eta}_Q
\end{equation}
where $\tilde{\mu}(\rho), \, \gamma, \, D_Q, \sigma_Q > 0$, $|\bm{Q}|^2 = \sum_{\alpha \beta} \bm{Q}^2_{\alpha \beta}$, 
$\widehat{\nabla\nabla} = \nabla \nabla - \frac{1}{2} \nabla^{2}$ and $\bm{\eta}_Q$ is a symmetric traceless tensor whose components 
are unit Gaussian white noises.
Upon replacing the activity $A$ with the density $\rho$ in the right hand side of Eq.~(\ref{eq:nematicevolution}), one recovers the standard dynamics for active nematics  \cite{ramaswamy2003,bertin2013}.

We now estimate density fluctuations deep in the nematically ordered phase, that is, for $\rho=\rho_0$ such that $\tilde{\mu}(\rho_0) = \mu_0 > 0$ (and not small), using a linearized theory.
Neglecting noises, Eqs.~(\ref{eq:densityevolution})--(\ref{eq:nematicevolution}) admit a set of homogeneous solutions with $A=(\kappa \rho_0 - a)/\lambda=A_0 \propto \Delta\phi$, $S = \sqrt{2 \mu_0/\gamma} = S_0$ and arbitrary nematic director $\bm{n}_0$ such that $\bm{Q}_0 = S_0(\bm{n}_0  \bm{n}_0- \frac{1}{2}\bm{1})$~\cite{bertin2013}.
We assume this homogeneous state to be linearly stable, which implicitly imposes conditions on parameter values.
In the presence of noise, we then consider small fluctuations of the density $\rho(\bm{r}, t) = \rho_0 + \delta \rho(\bm{r}, t)$, of the activity $A(\bm{r}, t) = A_0 + \delta A(\bm{r}, t)$, and of the nematic order $\bm{Q}(\bm{r}, t) = \bm{Q}_0 + \delta\bm{Q}(\bm{r}, t)$.
Linearizing Eqs.~(\ref{eq:densityevolution})--(\ref{eq:nematicevolution}) and introducing the spatial Fourier transform $\delta \hat{\rho}(\bm{q},t) = \int d\bm{r} \delta \rho (\bm{r}, t) \exp[-i \bm{r} \cdot \bm{q}]$ 
with $\bm{q} = q (\cos \theta \bm{n}_0 + \sin \theta \bm{n}_\perp)$ and $\bm{n}_\perp \cdot \bm{n}_0 = 0$,
one finds for the structure factor $S(\bm{q}) =  \langle \delta \hat{\rho}(\bm{q},t) \delta \hat{\rho}(-\bm{q}, t) \rangle$ the critical scaling behavior~\cite{SuppMat}
\begin{equation} \label{eq:Sq}
    \frac{S(\bm{q})}{\Delta\phi^{1/2}}  \propto  \frac{(1-\cos 4\theta) \lambda^2 \chi_2^2 \sigma_Q^2}{D_Q D_\rho \kappa (\kappa D_\rho + \lambda D_Q )} \frac{1}{\tilde{q}^2} \\ + \frac{2 D_\rho \sigma_A^2}{\kappa \lambda} \tilde{q}^2 \, ,
\end{equation}
where $\tilde{q} = q\xi$, with a crossover length $\xi=\Delta\phi^{-3/4}$.
In this scaling regime, we thus observe a crossover between GNF ($\lambda=-2$) for $q \ll q^* \sim \xi^{-1}$ and hyperuniformity ($\lambda = 2$) for $q \gg q^*$.
The predicted values $\zeta' = \frac{1}{2}$ and $\mu = \frac{3}{4}$ 
are reasonably close to the measured $\zeta' \approx 0.55$ and $\mu \approx 0.625$, meaning that the linearized theory already captures a significant part of the phenomenology.

An intuitive understanding of this crossover behavior may be gained as follows. In the large-length scale regime $q \ll A_0^{1/2}$,
the long-time scale dynamics of a density mode $\delta \hat{\rho}(\bm{q},t)$ may be approximated as \cite{SuppMat}
\begin{equation}\label{eq:lin_rho}
\partial_t \delta \hat{\rho} = -D q^2 \delta \hat{\rho} + \frac{\tilde{\sigma}_A q^2}{\sqrt{A_0}} \hat{\eta}_A + \tilde{\sigma}_Q \sin 2\theta A_0 \hat{\eta}_Q^{\perp},
\end{equation}
with $D=D_{\rho}\kappa/\lambda$, $\tilde{\sigma}_A = D_{\rho}\sigma_A/\lambda$, $\tilde{\sigma}_Q = \chi_2\sigma_Q/D_Q$,
and $\hat{\eta}_Q^{\perp} = \bm{n}_0 \cdot \hat{\bm{\eta}}_Q \cdot \bm{n}_{\perp}$.
We assumed for simplicity that $\chi_3=0$, as $\chi_3$ plays no role in Eq.~(\ref{eq:Sq}).
The dynamics of $\delta \hat{\rho}$ is thus a (complex) Langevin equation with two effective noise terms:
a `superconservative' noise of amplitude $\propto q^4/A_0$ induced by the activity field $A$,
and a seemingly nonconserved noise of amplitude $\propto A_0^2$ induced by the nematic field ${\bm Q}$.
Notably, the amplitudes of these two noises become comparable for $q^* \sim A_0^{3/4}$.
The conserved noise ($\hat{\bm{\eta}}_{\rho}$ term) of amplitude $\propto q^2 A_0$ obtained from linearizing Eq.~(\ref{eq:densityevolution}) is negligible in this regime. 
The noise $\hat{\eta}_A$ dominates over the noise $\hat{\eta}_Q^{\perp}$ for $q \gg q^*$ (intermediate scales) leading to the hyperuniform-like scaling $S(q) \sim q^2/A_0$.
The noise $\hat{\eta}_Q^{\perp}$ instead dominates for $q \ll q^*$ (larger scales), where one obtains $S(q) \sim (1-\cos 4\theta)A_0^2/q^2$.
Both results are in agreement with Eq.~(\ref{eq:Sq}) (recalling that $A_0 \sim \Delta\phi$) and allow us to rationalize, at a heuristic level, the critical crossover as originating from the competition of two effective noise sources, each dominating on different length scales.

In this work, we have shown that systems of nematic active particles with a diffusivity activated by the presence of neighboring particles display an absorbing phase transition with unusual properties. Close to the transition, 
density fluctuations are suppressed at intermediate length scales, but enhanced on large length scales. The crossover length 
diverges as a power law of the distance to the transition point, and is thus a critical property,
as also confirmed by a critical data collapse on both the number fluctuations and the structure factor.
We rationalize this behavior, at a qualitative level, by deriving the structure factor from a linearized continuum theory.
Our linearized theory captures the critical crossover between anomalous behaviors. At intermediate length scales, an effective noise, arising from the coupling with the activity field, controls and suppresses density fluctuations. At large length scales, density fluctuations are instead dominated by an effective noise arising from the coupling with the nematic tensor leading to GNF.
Interestingly, the behavior of density fluctuations is the opposite of the one observed in clustering or coarsening patterns in scalar active matter, where large density fluctuations occur on intermediate length scales, the system being hyperuniform on the largest length scales \cite{LowenPRR2024,zhangHyperuniformActiveChiral2022}.

Our work opens avenues for the investigation of absorbing phase transitions in orientationally-ordered active matter with environment-dependent motility, with experimental realizations ranging from \modif{suspensions of rods under cyclic shear \cite{franceschiniDynamicsNonBrownianFiber2014,trulssonDirectionalShearJamming2021} to} Quincke rollers with neighbor-triggered motility \cite{LiuActivityPNAS2021}, \modif{or} to microrobots that can sense their neighborhood and modify their behavior accordingly~\cite{muinos2021,benzion2023}. On the numerical side, future work may try to quantitatively characterize the anisotropic features of the structure factor, beyond angular averages.
On the theoretical side, whether the presence of orientational order modifies \modif{the activity-related critical exponents} 
of the absorbing phase transition of the ROM, which corresponds to Conserved
Directed Percolation (CDP) in the absence of orientational order \cite{hinrichsenNonequilibriumCriticalPhenomena2000}, remains an important open question,
which requires \modif{either extensive numerical simulations or} the development of renormalization schemes for continuum equations coupling activity, nematic (or polar) and particle density fields.

\nocite{henkes2020dense}
\textit{Acknowledgments:} 
SDC acknowledges support from the ANR-18-CE30-0028-01 grant LABS.


\bibliographystyle{apsrev4-2}
\bibliography{sample}

\end{document}


\title{Supplemental material for ``Giant density fluctuations in locally hyperuniform states''}

\author{Sara Dal Cengio}
\affiliation{Universit\'e Grenoble Alpes, CNRS, LIPhy, 38000 Grenoble, France}

\author{Romain Mari}
\affiliation{Universit\'e Grenoble Alpes, CNRS, LIPhy, 38000 Grenoble, France}

\author{Eric Bertin}
\affiliation{Universit\'e Grenoble Alpes, CNRS, LIPhy, 38000 Grenoble, France}

\maketitle


\section{Continuum theory}\label{app:meanfieldtheory}

\subsection{Evolution equations for the activity, density and nematic fields}

The starting point are the continuum equations for the three hydrodynamic fields the activity $A(\r, t)$, the particle density $\rho(\r, t)$ 
and the nematic tensor $\bm{Q}(\r, t) = S\left( \bm{n} \boldsymbol{n} - \frac{1}{2} \bm{1} \right)$, as they appear in the main text:
\begin{eqnarray}
    \partial_t A &=& \nabla \cdot \bm{J} + (\kappa \rho - a)  A - \lambda A^2 + \sigma_A \sqrt{A} \, \eta_A \label{eq:dynamicsA} \\
    \partial_t \rho &=& \nabla \cdot \bm{J} + \sigma_\rho \nabla \cdot \left( \sqrt{A} \bm{\eta}_{\rho}\right)  \label{eq:dynamicsrho}\\
     \partial_t \bm{Q} &=& \left(\tilde{\mu}(\rho) - \gamma |\bm{Q}|^{2}\right) \bm{Q} + D_{ Q} \nabla^2 \bm{Q} + \chi_3 \widehat{\nabla \nabla}  A  + \sigma_Q \bm{\eta}_Q \label{eq:dynamicsQ}
\end{eqnarray}
with $\bm{J} = D_{\rho} \nabla A + \chi_1 \bm{Q} \cdot \nabla A + \chi_2 A \nabla \cdot \bm{Q}$.
%
In the noiseless dynamics, the absorbing phase transition occurs at critical density $ \rho_0^{\rm c} = a/\kappa$ where the activity vanishes as $A \sim (\rho_0 - \rho_0^{\rm c})^{\beta}$ with mean-field exponent $\beta_{\rm CDP}^{\rm MF}=1$ larger than $\beta_{\rm CDP} \approx 0.64$  found in twodimensional numerical simulations \cite{Lubeck2004}.
%
In Eq.~(\ref{eq:dynamicsQ}), $|\bm{Q}|^2 = \sum_{\alpha \beta} \bm{Q}_{\alpha \beta} \bm{Q}_{\alpha \beta} = S^2 /2$ is the (square) Frobenius norm and $\widehat{\nabla \nabla}$ is the traceless symmetric differential operator $\widehat{\nabla \nabla} = \nabla \nabla - \frac{1}{2} \bm{1}\nabla^{2} $. 
In our model, both active and passive particles align with neighboring particles, but only the active particles transport the nematic tensor, which explains why the coupling in Eq.~(\ref{eq:dynamicsQ}) involves $A$ and not $\rho$. Neglecting the noise, Eq.~(\ref{eq:dynamicsQ}) admits a homogeneous nematically ordered solution  for $\mu_0 = \tilde{\mu}(\rho_0)  >0$ and $S = \sqrt{2 \mu_0/\gamma} = S_0$. In the following, we assume $\mu_0\gg 0$,  indicating that we are deep within the homogeneous nematic phase \cite{ramaswamy2003}
and denote $\bm{Q}_0 \equiv S_0 \left( \boldsymbol{n}_0 \boldsymbol{n}_0 - \frac{1}{2} \bm{1}\right)$ the homogeneous nematic tensor with uniaxial direction $\boldsymbol{n}_0 = \left( \cos \theta_0, \sin \theta_0 \right)$.

\subsection{Linearized equations}

To study small fluctuations about $\bm{Q}_0$, we consider $S = S_0 + \delta S(\r, t)$ and
 $\boldsymbol{n} (\r, t) = \boldsymbol{n}_0 +  \delta n_{\perp}(\r ,t) \;\boldsymbol{n}_{\perp}$, with $\boldsymbol{n}_{\perp} = \left(- \sin \theta_0, \cos \theta_0 \right)$ being the direction perpendicular to $\boldsymbol{n}_0$. We expand the nematic tensor accordingly:

\begin{equation}
    \bm{Q} = \bm{Q}_0 + \frac{1}{2}\begin{pmatrix}
        \cos 2 \theta_0 & \sin 2 \theta_0 \\
        \sin 2 \theta_0 & \cos 2 \theta_0
    \end{pmatrix} \, \delta S(\r, t)+ S_0
    \begin{pmatrix}
-\sin 2 \theta_0 & \cos 2 \theta_0 \\
 \cos 2 \theta_0 & \sin 2 \theta_0
\end{pmatrix} \, \delta n_{\perp} (\r, t) 
\end{equation}
and introduce new variables:
\begin{eqnarray}
\begin{cases}
    z_{\parallel} = \cos{\theta_0} x + \sin{\theta_0} y \\
    z_{\perp} = -\sin{\theta_0} x + \cos{\theta_0} y 
    \end{cases}
\end{eqnarray}
Notice that in the new reference frame the operator $\widehat{\nabla \nabla}$ becomes:
\begin{eqnarray}
    \widehat{\nabla \nabla}= \frac{1}{2} \begin{pmatrix}
        \cos{2 \theta_0} & - \sin{2 \theta_0} \\
        \sin{2 \theta_0} & \cos{2 \theta_0}
    \end{pmatrix} \begin{pmatrix}
        \partial_{\parallel}^2 - \partial_{\perp}^2 & 2 \partial_{\parallel}\partial_{\perp} \\
        2 \partial_{\parallel}\partial_{\perp} & \partial_{\perp}^{2} - \partial_{\parallel}^2 
    \end{pmatrix}
\end{eqnarray}
where we denote $\partial_\parallel$ and $\partial_\perp$ differentiation with respect to $z_{\parallel}$ and $z_{\perp}$, respectively.
Together with fluctuations in the nematic phase, we consider fluctuations in the homogeneous density profile $\rho(\r, t) = \rho_0 + \delta \rho(\r, t)$ and in the homogeneous  activity $A (\r, t) = A_0 + \delta A(\r ,t)$, with $A_0 \equiv \frac{\kappa}{\lambda}(\rho_0 - \rho_0^{\rm c})$.

All together, the linearized Equations ~(\ref{eq:dynamicsA})--(\ref{eq:dynamicsQ}) give:
\begin{eqnarray} 
    \partial_t \delta A &=& D_{\rho} \nabla^2 \delta A + S_0 \frac{\chi_1}{2} \left( \partial_\parallel^2 - \partial_\perp^2 \right) \delta A + A_0 \frac{\chi_2}{2} \left( \partial_\parallel^2 - \partial_\perp^2 \right) \delta S + 2  A_0 S_0 \chi_2\partial_{\parallel}\partial_{\perp} \delta n_{\perp} - \lambda A_0 \delta A +  \kappa A_0 \delta \rho + \sigma_A \sqrt{A_0} \eta_A \label{eq:linearizedA}\\
    \partial_t \delta \rho &=& D_{\rho} \nabla^2 \delta A + S_0 \frac{\chi_1}{2} (\partial_{\parallel}^2 - \partial_{\perp}^2) \delta A  +  A_0 \frac{\chi_2}{2} (\partial_{\parallel}^2 - \partial_{\perp}^2)\delta S + 2  A_0 S_0 \chi_2\partial_{\parallel}\partial_{\perp}\delta n_{\perp} + \sigma_\rho \sqrt{ A_0} \;\nabla \cdot  \boldsymbol{\eta}_{\rho} \label{eq:linearizedrho} \\
    S_0 \partial_t \delta n_{\perp} &=&  \Dq S_0 \nabla^2 \delta n_{\perp}+ \chi_3 \partial_{\parallel} \partial_{\perp} \delta A   + \sigma_{Q} \eta_{Q}^{\perp}\label{eq:linearizedQ} \\
    \frac{1}{2}\partial_t \delta S &=& \mu'_0 \frac{S_0}{2} \delta \rho - \gamma \frac{S_0^2}{2}\delta S + \frac{D_{\rm Q}}{2} \nabla^2 \delta S + \frac{\chi_3}{2} (\partial_{\parallel}^2 - \partial_{\perp}^2 ) \delta A  + \sigma_{Q} \eta_{Q}^{\parallel}\label{eq:linearizedS}
\end{eqnarray}
with  $ \eta_{Q}^{\parallel} = \bm{n}_0 \cdot \bm{\eta}_Q \cdot \bm{n}_0$, $ \eta_{Q}^{\perp} = \bm{n}_0 \cdot \bm{\eta}_Q \cdot \bm{n}_{\perp}$ and $\mu_0' = \left. \frac{d \tilde{\mu}}{d \rho} \right\vert_{\rho= \rho_0}$. We have retained only leading order terms $\propto \sqrt{A_0}$ in the multiplicative noises .
Taking the limit $\gamma \to \infty$ in Eq.~(\ref{eq:linearizedS}), we treat $\delta S$ as a fast mode of relaxation \cite{ramaswamy2003} and approximate it as $\delta S \approx \frac{\mu_0'}{\gamma S_0} \delta \rho +  \mathcal{O}(\nabla^2)$, treating the noise term as a higher order contribution. Doing so, we restrict our analysis to the slow modes $\delta A$, $\delta \rho$ and $\delta n_{\perp}$. Their corresponding evolution equations now read:
\begin{eqnarray}
    \partial_t \delta A &=& D_{\rho} \nabla^2 \delta A + S_0\frac{\chi_1}{2} \left( \partial_\parallel^2 - \partial_\perp^2 \right) \delta A + A_0 \frac{\chi_2 \mu_0'}{2 \gamma S_0} \left( \partial_\parallel^2 - \partial_\perp^2 \right) \delta \rho + 2 \chi_2 A_0 S_0 \partial_{\parallel}\partial_{\perp} \delta n_{\perp} - \lambda A_0 \delta A +  \kappa A_0 \delta \rho + \sigma_A \sqrt{A_0} \eta_A \label{eq:linearizedA2} \\
    \partial_t \delta \rho &=& D_{\rho} \nabla^2 \delta A + S_0 \frac{\chi_1}{2} (\partial_{\parallel}^2 - \partial_{\perp}^2) \delta A  +  A_0\frac{\chi_2 \mu_0'}{2 \gamma S_0} (\partial_{\parallel}^2 - \partial_{\perp}^2)\delta \rho + 2  A_0 S_0 \chi_2\partial_{\parallel}\partial_{\perp}\delta n_{\perp} + \sigma_\rho \sqrt{ A_0} \;\nabla \cdot  \boldsymbol{\eta}_{\rho} \label{eq:linearizedrho2} \\
     \partial_t \delta n_{\perp} &=&  \Dq \nabla^2 \delta n_{\perp}+ \frac{\chi_3}{S_0} \partial_{\parallel} \partial_{\perp} \delta A   + \frac{\sigma_{Q}}{S_0} \eta_{Q}^{\perp}\label{eq:linearizedQ2}
\end{eqnarray}

\subsection{Spatiotemporal Fourier analysis}

Eqs.~(\ref{eq:linearizedA2})--(\ref{eq:linearizedQ2}) are most conveniently analysed in Fourier space, using the time-and-space Fourier transform $\tilde{\mathcal{F}}(\q, \omega) = \int dt \int d\r \mathcal{F} (\r, t) \exp[-i \r \cdot \q + i \omega t]$ :
\begin{align} 
    -i \omega \delta \tilde{A}(\q, \omega) &= - \left( q^2 D_{\rho} + q^2  S_0 \frac{\chi_1}{2} \cos{2\theta} + \lambda A_0 \right) \delta\tilde{A} - \left(\frac{A_0 \chi_2 \mu_0'}{2 \gamma S_0}q^2 \cos{2 \theta} - \kappa A_0 \right) \delta \tilde{\rho} - \chi_2 A_0 S_0 q^2 \sin{2 \theta} \delta \tilde{n}_{\perp} + \sigma_A \sqrt{A_0} \tilde{\eta}_A \label{eq:fourierA} \\
    -i \omega \delta \tilde{\rho} (\q, \omega) &= - q^2 \left(D_{\rho} + S_0 \frac{\chi_1}{2} \cos{2\theta} \right) \delta\tilde{A} -\frac{A_0 \chi_2 \mu_0'}{2 \gamma S_0}q^2 \cos{2 \theta} \delta \tilde{\rho} - \chi_2 A_0 S_0 q^2 \sin{2 \theta} \delta \tilde{n}_{\perp} + i q \sigma_{\rho} \sqrt{A_0} \bm{\hat{q}} \cdot \boldsymbol{\tilde{\eta}}_{\rho}\label{eq:fourierrho} \\
    -i \omega \delta \tilde{n}_{\perp}(\q, \omega) &=- q^2 \Dq \delta \tilde{n}_{\perp} -  q^2 \frac{\chi_3}{2 S_0} \sin{2 \theta} \delta \tilde{A}   + \frac{\sigma_{Q}}{S_0}\tilde{\eta}_{Q}^{\perp}  \label{eq:fourierQ}
\end{align}
with $\bm{q} = q (\cos{\theta} \bm{n}_0 + \sin{\theta} \bm{n}_{\perp}) = q \bm{\hat{q}} $,
%
In matricial form, Eqs.~(\ref{eq:fourierA})--(\ref{eq:fourierQ}) can be written as:
\begin{equation}
\left( - i \omega \bm{1} - \bm{M} \right) \boldsymbol{\delta S} = \boldsymbol{b}
\end{equation}
with $\boldsymbol{\delta S} = \left( \delta \tilde{A}\, ,  \, \delta \tilde{\rho} \, , \,  \delta \tilde{n}_{\perp} \right)^{\top}$, $\boldsymbol{b} = \left( \sigma_A \sqrt{A_0} \tilde{\eta}_A \, , \, i  q  \sigma_{\rho} \sqrt{A_0} \boldsymbol{\hat{q}}\cdot \boldsymbol{\tilde{\eta}}_{\rho}\, , \, \sigma_{Q} \tilde{\eta}^{\perp}_{Q}/S_0 \right)^{\top} $, and $\bm{M}$ being the matrix of linear stability:
\begin{equation}\label{eq:matrixlinearstability}
     \bm{M} = \begin{pmatrix}
        - q^2 \mathcal{D}_{\rho}(\theta) - \lambda A_0 & \kappa A_0 - \frac{A_0 \chi_2 \mu_0'}{2 \gamma S_0}q^2 \cos{2 \theta} & -\chi_2 A_0 S_0 q^2 \sin{2 \theta}  \\
        - q^2\mathcal{D}_{\rho}(\theta) & - \frac{A_0 \chi_2 \mu_0'}{2 \gamma S_0}q^2 \cos{2 \theta} & -\chi_2 A_0 S_0 q^2 \sin{2 \theta} \\
         - q^2 \frac{\chi_3}{2 S_0} \sin{2 \theta} & 0   & - q^2 \Dq 
    \end{pmatrix}
\end{equation}
with $\mathcal{D}_{\rho}(\theta)= D_{\rho} + S_0\frac{\chi_1}{2} \cos{2\theta} $.
Using Cramer's formula we can explicitly express the components of the vector $\boldsymbol{\delta S}$ as:
\begin{equation}\label{eq:cramer}
    \boldsymbol{\delta S}_i = \frac{\det \left( - i \omega \bm{1} - \bm{M}\right)\vert_{\boldsymbol{i} \to \boldsymbol{b}} }{\det (- i \omega \bm{1} - \bm{M})}
\end{equation}
where, in the numerator, the $i$th-column of the matrix $( - i \omega \bm{1} - \bm{M})$ is replaced with the vector $\boldsymbol{b}$. We are interested in the number density fluctuations $\delta \rho$ which, using Eq.~(\ref{eq:cramer}), are controlled by the determinant of the following matrix: 
\begin{eqnarray}
 \det \begin{pmatrix}
        q^2 \mathcal{D}_{\rho}(\theta) + \lambda A_0 - i \omega &  \sigma_A\sqrt{A_0} \tilde{\eta}_{A}
 & \chi_2 A_0 S_0 q^2 \sin{2 \theta} \\
   q^2\mathcal{D}_{\rho}(\theta) & i q \sigma_{\rho}\sqrt{A_0}  \boldsymbol{\hat{q}} \cdot \boldsymbol{\tilde{\eta}}_{\rho} & \chi_2 A_0 S_0 q^2 \sin{2 \theta} \nonumber \\
  q^2 \frac{\chi_3}{2 S_0} \sin{2 \theta}& \sigma_{Q} \frac{\tilde{\eta}_Q^{\perp}}{S_0} & \Dq q^2 - i \omega 
 \end{pmatrix}
\end{eqnarray}
Notice that the denominator in Eq.~(\ref{eq:cramer}) is essentially the characteristic polynomial of $\bm{M}$ expressed in the variable $(- i \omega)$. We denote the eigenvalues of  $\bm{M}$ by $\mu_{1, 2, 3} (\boldsymbol{q})$ and assume linear stability, i.e. $\operatorname{Re}[\mu_{1, 2, 3}(\boldsymbol{q})] <0$. 
Accordingly, the Fourier mode $\delta \rho(\bm{q},\omega)$ of the density fluctuations reads:
\begin{eqnarray}\label{eq:solutiondeltarho}
    \delta \rho(\bm{q},\omega) &=& \frac{\sigma_A \sqrt{A_0} q^2 \left[2 (D_Q q^2 - i \omega) \mathcal{D}_{\rho}(\theta)- A_0 q^2 X_{23}(\theta)\right]}{2 (i \omega + \mu_1)(i \omega + \mu_2)(i \omega + \mu_3)} \tilde{\eta}_{A}(\bm{q},\omega) + \\ &-& \frac{i q \sqrt{A_0} \sigma_{\rho} \left[ 2(D_Q q^2 - i \omega) (\mathcal{D}_{\rho}(\theta)q^2 +  A_0 \lambda -  i \omega) - A_0 q^4 X_{23}(\theta)\right]}{2(i \omega + \mu_1)(i \omega + \mu_2)(i \omega + \mu_3)} \bm{\hat{q}} \cdot \bm{\tilde{\eta}}_{\rho}(\bm{q},\omega)  + \\ &+& \frac{\sigma_Q A_0 q^2 \chi_2 (A_0 \lambda - i \omega) \sin{2 \theta}}{(i \omega + \mu_1)(i \omega + \mu_2)(i \omega + \mu_3)}\tilde{\eta}_{Q}(\bm{q},\omega)
\end{eqnarray}
with $X_{23}(\theta) = \chi_2 \chi_3 \sin^2{2 \theta}$.

\subsection{Structure factor}

\begin{figure}
\centering
\includegraphics[width=0.1\textwidth,angle=270]{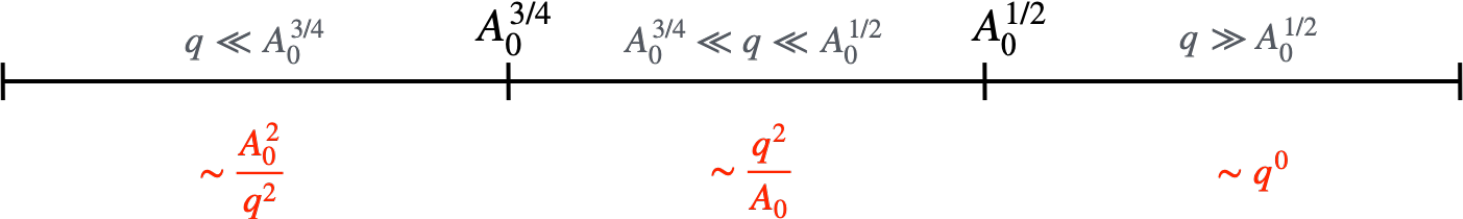}
\caption{
Sketch of the different ranges for  $q$ which corresponds to different scaling behavior for the structure factor (in red). }
\label{fig:figure1app}
\end{figure}

We now aim at determining the structure factor $S(\bm{q})=N^{-1}\langle \delta \rho(\q, t) \delta \rho^* (\q', t) \rangle$.
We first note that all fields on the right hand side of Eq.~(\ref{eq:solutiondeltarho}) have known autocorrelations. For instance,
\begin{eqnarray}
\langle \tilde{\eta}_A(\boldsymbol{q}, \omega) \tilde{\eta}_A(\boldsymbol{q}', \omega') \rangle &=& \int dt dt' d\r d\r' \langle \eta_A( \r, t) \eta_A (\r', t') \rangle e^{+i \omega t + i \omega' t' - i \r \cdot \q - i \r' \cdot \q'} \\
&=& \int dt d\r e^{i (\omega + \omega')t } e^{-i \r \cdot (\q + \q')} \\
&=& (2 \pi)^3 \delta (\omega + \omega') \delta (\q + \q') \, , 
\end{eqnarray}
and this similarly holds for $\tilde{\eta}_{Q}(\boldsymbol{q}, \omega)$ and $\boldsymbol{\tilde{\eta}}_{\rho}(\boldsymbol{q}, \omega)$. 
Thus, autocorrelating Eq.~(\ref{eq:solutiondeltarho}) gives:
\begin{eqnarray}\label{eq:powerspectrum}
    &&\langle \delta \rho (\boldsymbol{q},  \omega) \delta \rho^* (\boldsymbol{q'}, \omega') \rangle = \Big[ \frac{A_0 q^4 \sigma_A^2 \left( q^4 (2 D_Q \mathcal{D}_{\rho}(\theta) - A_0 X_{23}(\theta))^2 + 4 \mathcal{D}_{\rho}(\theta)^2 \omega^2 \right)}{4 (\omega^2 + \mu_1^2)(\omega^2 + \mu_2^2)(\omega^2 + \mu_3^2)}] + \nonumber \\ &+& \frac{A_0 q^2 \sigma_{\rho}^2 \left( q^4 (2 D_Q (\mathcal{D}_{\rho}(\theta) q^2 + A_0 \lambda) - A_0 q^2 X_{23}(\theta))^2 + 4 ( D_Q^2 q^4 + (\mathcal{D}_{\rho}(\theta)q^2 + A_0 \lambda)^2 + A_0 q^4 X_{23}(\theta)) \omega^2 + 4 \omega^4 \right)}{4 (\omega^2 + \mu_1^2)(\omega^2 + \mu_2^2)(\omega^2 + \mu_3^2)} + \nonumber \\ &+& \frac{A_0^2 q^4 \sigma_Q^2 \chi_2^2 \left( A_0^2 \lambda^2 + \omega^2\right) \sin^2{2 \theta}}{ (\omega^2 + \mu_1^2)(\omega^2 + \mu_2^2)(\omega^2 + \mu_3^2)} \Big] (2 \pi)^3 \delta(\omega+\omega') \delta (\bm{q}+ \bm{q'})
\end{eqnarray}
where the star denotes the complex conjugate. The equal-time Fourier transform (which is proportional to the structure factor $S(\bm{q})$) is obtained by integrating over the frequency, namely $
    \langle \delta \rho(\q, t) \delta \rho^* (\q', t) \rangle = \frac{1}{(2 \pi)^2} \int d\omega \int d\omega' \langle \delta \rho (\q, \omega) \delta \rho^* (\q', \omega') \rangle e^{-i (\omega + \omega') t} 
   $ which, using Eq.~(\ref{eq:powerspectrum}), gives:

\begin{eqnarray}\label{eq:solutionSq}
\langle \delta \rho(\q, t) \delta \rho^* (\q', t) \rangle = 4 \pi^2 \delta (\q + \q') \Bigg[ \frac{C + \frac{q^2}{A_0}  \mathcal{P}_1(A_0) + \frac{q^4}{A_0^3} \mathcal{P}_2(A_0) + \frac{q^6}{A_0^4}\mathcal{P}_3(A_0) + \frac{q^8}{A_0^4} \mathcal{P}_4(A_0) }{ \frac{q^2}{A_0^2} \mathcal{P}_5(A_0) + \frac{q^4}{A_0^3}\mathcal{P}_6(A_0) + \frac{q^6}{A_0^4}\mathcal{P}_7(A_0) }\Bigg]
\end{eqnarray}
where $C = 4 S_0^3 \lambda^3 \sigma_{Q}^2 \chi_2^2 \sin^2{2\theta}$ and the polynomials $\mathcal{P}_i(A_0)$ with $i = 1 \dots 7$ are function of $A_0$ independent on $q$ which satisfy $\mathcal{P}_i(0) \neq 0 \forall i$. Notice
 that $\delta(\q + \q')$ diverges when $\q' = - \q$ and, for a system of finite size $L$, is replaced with a Kronecker delta according to $\delta( \q + \q') \to \delta_{\q', -\q} \frac{L^2}{(2 \pi)^2}$ \cite{henkes2020dense}. 
 Eq.~(\ref{eq:solutionSq}) then relates directly to the structure factor $S(\q) = \frac{1}{N}\langle \delta \rho(\q, t) \delta \rho^* (-\q, t) \rangle$.
 We now introduce a rescaled variable $\Tilde{q} \equiv \frac{q}{A_0^\ex}$ so that:
\begin{eqnarray}\label{eq:scalingstructurefactor1}
    S(\q) &\propto& \frac{C + \Tilde{q}^2 A_0^{2 \ex -1} \mathcal{P}_1(A_0)+\Tilde{q}^4 A_0^{4 \ex -3} \mathcal{P}_2(A_0)+ \Tilde{q}^6 A_0^{6 \ex -4} \mathcal{P}_3(A_0)+ \Tilde{q}^8 A_0^{8 \ex -4} \mathcal{P}_4(A_0) }{\Tilde{q}^2 A_0^{2 \ex -2} \mathcal{P}_5(A_0)+\Tilde{q}^4 A_0^{4 \ex -3} \mathcal{P}_6(A_0)+ \Tilde{q}^6 A_0^{6 \ex -4} \mathcal{P}_7(A_0)} \, \nonumber \\
    &=& \frac{C A_0^{2 - 2 \ex} + \Tilde{q}^2 A_0 \mathcal{P}_1(A_0) + \Tilde{q}^4 A_0^{2 \mu -1} \mathcal{P}_2(A_0) + \Tilde{q}^6 A_0^{4 \mu-2} \mathcal{P}_3(A_0)+ \Tilde{q}^8 A_0^{6 \ex -2} \mathcal{P}_4(A_0)}{\Tilde{q}^2 \mathcal{P}_5(A_0) + \Tilde{q}^4 A_0^{2 \ex -1} \mathcal{P}_6(A_0)+ \Tilde{q}^6 A_0^{4 \ex -2} \mathcal{P}_7(A_0)}
\end{eqnarray}
where in the second line we have factorized the first term in the denominator which diverges in the limit $ A_0 \to 0$ when $\mu <1$. From Eq.~(\ref{eq:scalingstructurefactor1}) we identify two exponents  $\ex = \frac{1}{2}$ and $\ex = \frac{3}{4}$ for which the structure factor exhibits a nontrivial scaling relation with a crossover in the $A_0 \to 0$ limit, (See Fig.~\ref{fig:figure1app}). For $\mu = \frac{1}{2}$ the scaling relation when $A_0 \to 0$ is:
\begin{equation}
 S(\q) \propto \frac{\Tilde{q}^2 \mathcal{P}_2 (0) + \Tilde{q}^4 \mathcal{P}_3(0) }{\mathcal{P}_5 (0) + \Tilde{q}^2 \mathcal{P}_6 (0) + \Tilde{q}^4 \mathcal{P}_7 (0)}
\end{equation}
with a crossover between normal behavior $S(\q) \sim q^0$ for $\Tilde{q} \gg 1$ and hyperuniform behavior $S(\q) \sim \Tilde{q}^2$ for $\Tilde{q} \ll 1$. More interestingly, for $\mu= \frac{3}{4}$ we get:
\begin{eqnarray}\label{eq:crossover}
    S(\q) \propto A_0^{1/2}\left(\frac{C}{\Tilde{q}^2 \mathcal{P}_5(0)} + \Tilde{q}^2 \frac{\mathcal{P}_2(0)}{\mathcal{P}_5(0)}\right)
\end{eqnarray}
with a crossover between hyperuniformity  $S(\q) \sim q^2$ for $\Tilde{q} \gg 1$ and giant number fluctuations $S(\q) \sim q^{-2}$ for $\Tilde{q} \ll 1$. This crossover occurs at larger lengthscale than the previous one and it is consistent with the crossover captured in our simulations. Replacing the explicit expressions for the polynomials in Eq.~(\ref{eq:crossover}) and using the fact that $A_0 \sim \Delta \phi $, one gets:
\begin{equation}
    \frac{S(\q)}{ \Delta \phi ^{1/2}}\propto \frac{(1 - \cos{4 \theta}) \lambda^2 \sigma_{Q}^2 \chi_2^2}{\Dq  \mathcal{D}_{\rho}(\theta)^2 \kappa( \kappa \mathcal{D}_{\rho}(\theta) + D_Q \lambda)} + \frac{ 2\sigma_A^2 \mathcal{D}_{\rho}(\theta)}{ \kappa \lambda} \Tilde{q}^2
\end{equation}
We recall that by taking $\chi_1 =0$, $\mathcal{D}_{\rho}(\theta)$ reduces to the parameter $D_{\rho}$ and one recover the formula in the main text.

\subsection{Approximate large-time Langevin equation for $\delta\rho(q,t)$}

To obtain the simplified Langevin equation for $\delta\hat{\rho}(q,t)$ given in Eq.~(8) of the main text, we use several assumptions and approximations.
First, we stick to the case $\chi_1=0$ considered in the main text, and we further assume that $\chi_3=0$ to simplify calculations, since the detailed calculation of the structure factor $S(q)$ shows that $\chi_3$ plays no role at leading order in the regime considered. Second, and in line with calculations done in \cite{ramaswamy2003}, we neglect the fluctuations $\delta S$ of the amplitude of nematic order, and only consider fluctuations $\delta n_{\perp}$ of the orientation $\bm{n}$. Formally, the corresponds to taking the limit $\gamma\to \infty$ in Eq.~(\ref{eq:fourierrho}).
Third, we make the approximation that on large enough time scales, $\omega$ can be neglected with respect to $q^2 D_\rho$ and $q^2 D_Q$ in Eqs.~(\ref{eq:fourierA}) and (\ref{eq:fourierQ}).
Finally, we assume that $q\ll A_0^{1/2}$ and we get
\begin{equation}
    \delta \tilde{A} \approx \frac{\kappa}{\lambda} \delta\tilde{\rho} + \frac{\sigma_A}{\lambda A_0^{1/2}} \tilde{\eta}_A, \qquad
    \delta \tilde{n}_{\perp} \approx \frac{\sigma_Q}{S_0 D_Q q^2} \tilde{\eta}_Q^{\perp}
\end{equation}
Injecting these expressions of $\delta \tilde{A}$ and $\delta \tilde{n}_{\perp}$ in Eq.~(\ref{eq:fourierrho}) then yields Eq.~(8) of the main text, once written in real time
(we did not write the conservative noise proportional to $\bm{\eta}_{\rho}$ because it is negligible in the regime considered, as argued in the main text).
Quite importantly, the obtained Langevin equation is only a heuristic description that applies on long time scales, due to the approximations made and the restricted range of frequency $\omega$ over which it is supposed to be (approximately) valid. However, this heuristic description already leads to the correct scaling of the crossover length scale $\xi$ (or equivalently, of $q^*\sim \xi^{-1}$) and of the structure factor in both regimes. Even the angular dependence of the structure factor in the regime $q\ll q^*$ is correctly reproduced. Hence this simple argument already reproduces the key physics of the problem, and provides an intuitive interpretation of the crossover observed in the structure factor as a competition between two effective noises resulting from the activity and nematic fields respectively, with different $A_0$ and $q$ dependences. In this way, it also highlights the key role played by a three-field
theory in obtaining a crossover between two different types of anomalous fluctuations as a function of scale.

\section{Time evolution of number fluctuations}
\begin{figure}[h!]
\centering
\includegraphics[width=0.5\linewidth]{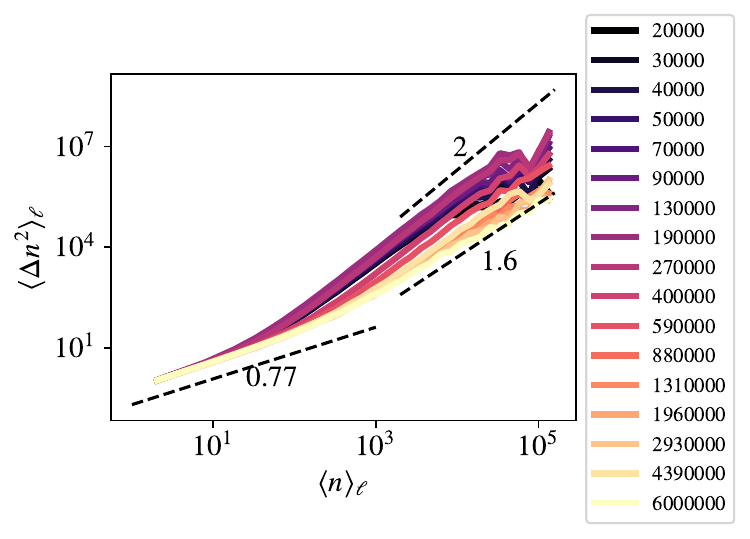}
\caption{Number fluctuations $\langle \Delta n^2  \rangle_{\ell}= \langle n^2 \rangle_{\ell} - \langle n \rangle_{\ell}^2$ versus $\langle n \rangle_{\ell}$ in logarithmic scale for $\phi =0.34$ at different timesteps. }
\label{fig:timeseriesnbfluctuation}
\end{figure}

In Fig.~\ref{fig:timeseriesnbfluctuation}, we report the variance
of particle number  as a function of the average particle number over consecutive time steps. Here $\phi=0.34$, $L=516$, $\delta=0.3$, $R_{\rm al} =1$ and $D=0.46$. The simulation is initialized with random angles and random positions and it evolves for $t_{\max} =6 \times10^6$ timesteps. The figure illustrates the onset of hyperuniformity at smaller length scale and giant number fluctuations at larger length scale during the relaxation dynamics. Interestingly, at large length scale, the system goes through an overshoot with an exponent close to $2$ (typical of phase separated systems) before stabilizing around $1.6$, a value consistent with the one reported in the literature of active nematics \cite{ngo2014}. This overshoot reflects transient inhomogeneities in the density field, which correlate with the presence of nematic defects. Over time, these defects annihilate, leading to a homogeneous state with global nematic order.

\begin{figure}[h!]
    \centering
    \begin{minipage}{0.49\linewidth}
        \centering
        \includegraphics[width=\linewidth]{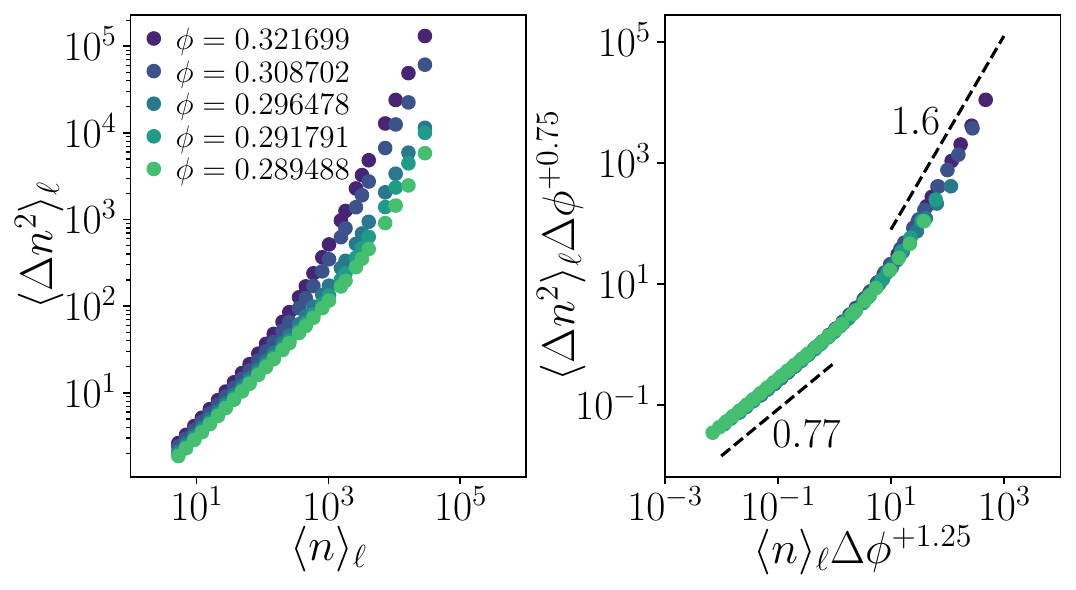}
    \end{minipage}
    \begin{minipage}{0.49\linewidth}
        \centering
        \includegraphics[width=\linewidth]{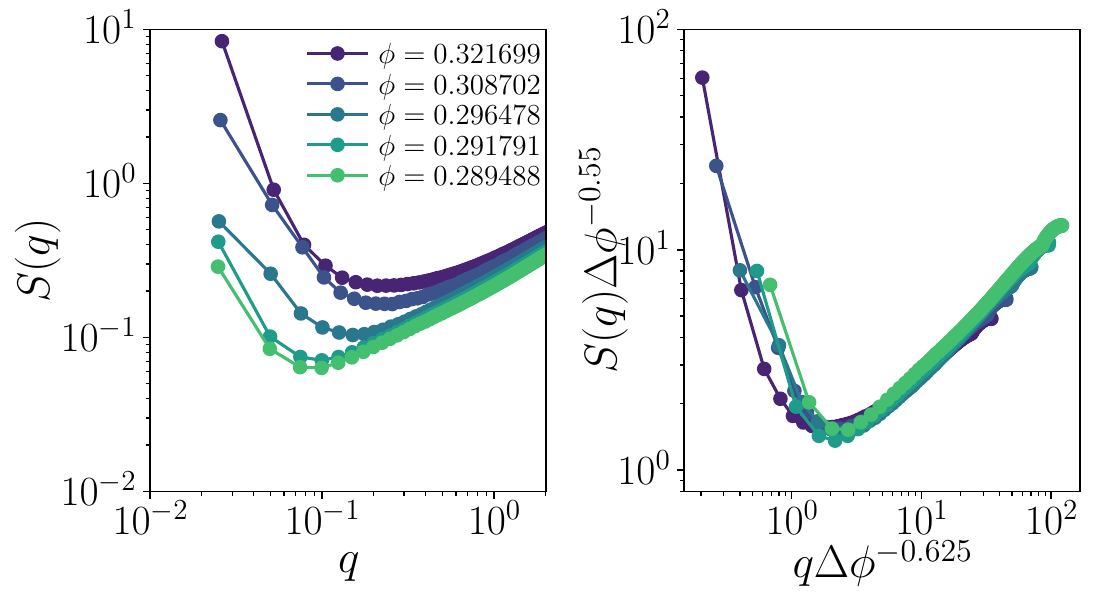}
    \end{minipage}
    \caption{From left to right: Number fluctuations $\langle \Delta n^2  \rangle_{\ell}= \langle n^2 \rangle_{\ell} - \langle n \rangle_{\ell}^2$ versus $\langle n \rangle_{\ell}$ in logarithmic scale for several packing fractions $\phi > \phi_\mathrm{c} = 0.2846$; Same data, rescaled as $\langle \Delta n^2 \rangle_{\ell} / \Delta \phi^{-0.75}$ versus $\langle n \rangle_{\ell}/\Delta \phi^{-1.25}$; Structure factor $S(q)$ for the same values of $\phi > \phi_\mathrm{c}$ in logarithmic scale; Same data, rescaled as $S(q)/\Delta \phi^{0.55}$ versus $q/\Delta \phi^{0.625}$.}
    \label{fig:rho4}
\end{figure}
\section{A different choice of parameters}
In Fig.~\ref{fig:rho4}, we present numerical results for a different set of parameters than those reported in the main text: $\sigma=0.05$, $\delta = 0.3$, $R_{\rm al} = 1$ and $D=0.3$. For this choice, the critical packing fraction is $\phi_c = 0.2846$,  which corresponds to a density $\rho \simeq 4 $. The results are in quantitative agreement with those reported in the main text: intermediate length scales display hyperuniformity, while large length scales show giant number fluctuations.
Furthermore the crossover is governed by the same critical exponents as reported in the main text for both number fluctuations and structure factor.

\section{Structure factor in the absorbing phase}

\begin{figure}
\centering
\includegraphics[width=0.9\textwidth]{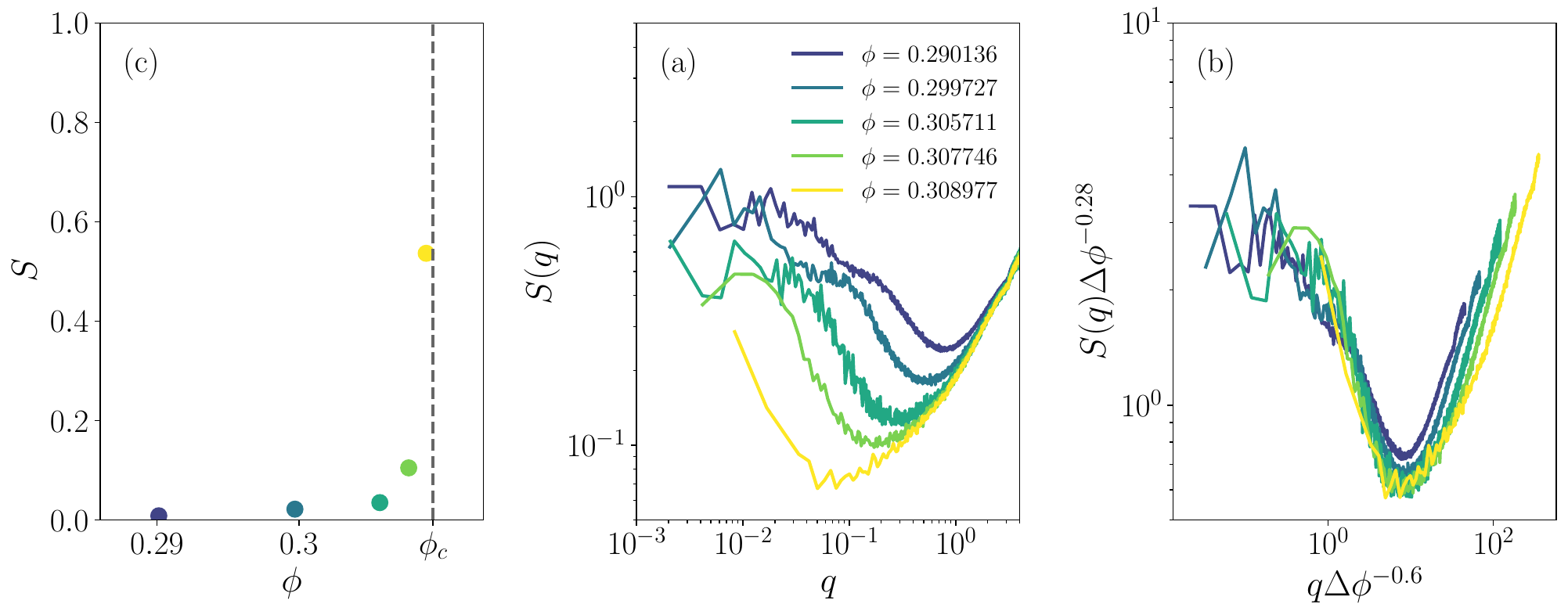}
\caption{
(a) Structure factor $S(q)$ for several values of $\phi<\phi_c$, with the same parameter values as for Fig. 2 in the main text, $\sigma = 0.1$, $\delta_0 = 0.3$, $R_\mathrm{al} = 1$, and $D = 0.46$. (b) Same data, rescaled as $S(q)/\Delta\phi^{0.28}$ as a function of $q/\Delta\phi^{0.6}$.}
\label{fig:structure_factor_below_phic}
\end{figure}

In this section, we show the progressive build up of hyperuniformity when approaching the absorbing phase transition from below, that is, for $\phi\to \phi_c^-$. 
In this phase, the system falls in an absorbing state in a finite time. 
As a consequence, neither hyperuniformity nor (and perhaps more importantly) giant number fluctuations have time to fully develop.
We report in Fig.~\ref{fig:structure_factor_below_phic}(a) the typical values of $S$ in the absorbing state for several values of $\phi<\phi_c$, and for the same parameter values than in the main text, that is, $\sigma = 0.1$, $\delta_0 = 0.3$, $R_\mathrm{al} = 1$, and $D = 0.46$. 
Even close to the transition, we find $S$ significantly smaller than $1$.

The structure factors measured in the absorbing state are shown in panel (b). 
The consequence of having no proper nematic ordering by the time the system falls into an absorbing state is that the behavior on this side of the transition is reminiscent of the one observed in the usual ROM~\cite{hexner2015,hexnerNoiseDiffusionHyperuniformity2017a}. At very small $q$ we observe normal fluctuations, then a minimum when increasing $q$ and then a remnant of hyperuniformity for larger $q$. 
Crucially, for $\phi<\phi_c$, $S(q)$ does not display signals of giant number fluctuations in the small $q$ limit where it instead displays normal fluctuations.

In panel (c), we show how the $\phi$ dependence can be partially captured by a rescaling $S(q)/\Delta\phi^{0.28}$ as a function of $q/\Delta\phi^{0.6}$. For comparison, for the usual ROM it is reported that for $\phi<\phi_c$, collapse is achieved with $S(q)\Delta\phi^{-0.44}$ as a function of $q\Delta\phi^{-0.8}$~\cite{hexner2015}.

\bibliographystyle{ieeetr}
\bibliography{sample}